\documentclass[aps,twocolumn,prl,tightenlines,floatfix,showpacs]{revtex4}
\usepackage[dvips]{graphicx}
\usepackage[english]{babel}
\usepackage{amsmath}
\usepackage{amssymb}
\usepackage{times}
\newcommand{\bs}{\begin{split}}
\newcommand{\es}{\end{split}}
\usepackage[dvips]{graphicx}
\newcommand{\be}{\begin{equation}}
\newcommand{\ee}{\end{equation}}
\newcommand{\ba}{\begin{eqnarray}}
\newcommand{\ea}{\end{eqnarray}}

\newcommand{\Ek}{E_{\mathbf{k}}}

\newcommand{\xik}{\xi_{\mathbf{k}}}
\newcommand{\sumk}{\sum_{\mathbf{k}}}

\newcommand{\sumq}{\sum_{\mathbf{q}}}

\newcommand{\Omegaq}{\Omega_{\mathbf{q}}}

\newcommand{\p}{\partial}

\begin{document}

\title{First and second sound modes at finite temperature in
trapped Fermi gases from BCS to BEC}

\author{Yan He, Qijin Chen, Chih-Chun Chien, and K. Levin}

\affiliation{James Franck Institute and Department of Physics,
University of Chicago, Chicago, Illinois 60637}

\date{\today}

\begin{abstract}
  We determine the temperature $T$ dependence of first and second sound
  mode frequencies for trapped Fermi gases undergoing BCS to Bose
  Einstein condensation (BEC) crossover.  Our results are based on the
  two fluid equations in conjunction with a microscopic calculation of
  thermodynamical variables. As in experiment and at unitarity, we show
  that the lowest radial breathing mode is $T$ independent.  At finite
  $T$, higher order breathing modes strongly mix with second sound.
  Their complex $T$ dependence should provide an alternative way of
  measuring the transition temperature, $T_c$.
\end{abstract}

\pacs{03.75.Hh, 03.75.Ss, 74.20.-z \hfill \textbf{\textsf{arXiv:0704.1889}} }

\maketitle

The recent discovery of the superfluid phases of trapped Fermi gases has
led to considerable interest in their collective mode spectrum
\cite{Grimm4,Thomas2,Kinast2,GrimmCollective,Thomasnew,ThomasUnitary,ThomasSound}.
Among the modes of experimental interest are breathing modes as well as
propagating first sound.  While originally theoretical attention
\cite{Tosi,Heiselberg,MCCollective} was focused on ground state
properties experimental measurements are naturally not restricted to
temperature $T=0$.  Indeed there is an interesting body of information
which is emerging in these Fermi gases about the finite temperature
behavior \cite{Thomasnew,ThomasUnitary,GrimmCollective} of the breathing
modes and, more recently about the propagating sound velocity
\cite{ThomasSound}.

The purpose of this paper is to compute sound mode frequencies in
spherically trapped Fermi gases undergoing BCS to Bose-Einstein
condensation (BEC) crossover, at general $T$.  We present a solution of
the linearized two fluid equations and compare with recent experiments.
We focus on the radial breathing modes and present predictions for
second sound, as well.  The structure of the two fluid equations for
Bose \cite{GriffinZaremba} and Fermi gases \cite{TaylorGriffin} has been
rather extensively discussed.  In the crossover regime, the normal fluid
is novel \cite{ChenPRL98,ourreview}, containing both fermions and
noncondensed pairs, which have not been systematically included in
previous collective mode literature.

Of great importance to the field as a whole is the future possibility of
second sound observations, in large part because this may ultimately
help assign more precise experimental values to the transition
temperature $T_c$. While existing experimental techniques such as vortex
observation \cite{KetterleV} and density profile features \cite{ZSSK06}
help establish superfluidity they provide lower bounds on $T_c$ or
determine its value for the special case of a population imbalanced
system.  Thermodynamical experiments measure $T_c$ more directly
\cite{ThermoScience,ThomasEntropy} but have been confined to unitarity.
Thus other techniques, such as second sound observation will be of great
value.  One of the principal results of the present paper is an analysis
of how second sound is coupled to the breathing modes.  We demonstrate
that higher order breathing modes will reveal $T_c$ through this
coupling, and therefore are an alternative to direct second sound
measurements.  However, the lowest breathing mode appears to be
remarkably $T$ independent at unitarity. This has been observed
experimentally \cite{Thomasnew,GrimmCollective} and argued to follow
from isentropic considerations \cite{ThomasUnitary}. Here we show, that
even when we treat the full coupling between first and second sound, we
obtain similar $T$ independent behavior at unitarity.

At the core of the two fluid theory is the assumption that hydrodynamics
is valid and that there are frequent collisions which produce a state of
local thermodynamical equilibrium.  Although there are some exceptions
\cite{Stamper-Kurn}, reaching the two fluid regime has not been easy for
atomic Bose gases. Two fluid dynamics are more readily achieved for
Fermi gases, principally because in the crossover regime the large
scattering lengths produce sufficient collisions.  Nevertheless there
has been considerable theoretical interest in setting up
\cite{TaylorGriffin} and solving \cite{HoShenoy} the two fluid equations
for Bose condensates.  Indeed, hydrodynamical approaches have
successfully addressed both the $T=0$ and normal state regimes of the
Bose gases \cite{StringariGriffin} and demonstrated that the breathing
mode frequencies are the same.  Here, by contrast we address the Fermi
gas case in a trap.  Because they interact more strongly near unitarity,
hydrodynamical descriptions have been argued quite convincingly
\cite{ThomasUnitary,Heiselberg,Heiselberg5} to be valid.

Previous theoretical work has been confined to $T=0$ treatments of a
harmonic trap, or to finite $T$ theories \cite{Heiselberg5} of a uniform
gas.  Our work is most similar in spirit to an earlier Bose gas study
\cite{HoShenoy} although we introduce different numerical techniques as
well as address fermions rather than bosons.  We note that the input
thermodynamics of systems undergoing the BCS-BEC crossover which is used
in the present paper, has been rather well calibrated against
experimental measurements in Ref.~\cite{ThermoScience} and is based on a
finite temperature extension of the simplest (BCS-Leggett) ground state.
In the absence of a trap our results are for the most part similar to
those in Ref.~\cite{Heiselberg5}.

We begin with the two fluid equations which describe the dynamical
coupling of the superfluid velocity $\mathbf{v}_{s}$ and the normal
velocity $\mathbf{v}_{n}$.  Just as in the spirit of the original Landau
two fluid equations we ignore viscosity terms.  In the presence of a
trap potential $V_{ext}=\frac{1}{2} m\omega_{ho} r^2$, the two fluid
equations are given by $m\frac{\partial \mathbf{v}_{s}}{\partial
  t}+\nabla(\mu+V_{ext}+m \frac{ \mathbf{v}_{s}^{2}}{2})=0 $,
$\frac{\partial \mathbf{j}}{\partial t}+\nabla \cdot \mathbf{\Pi} =-n
\nabla V_{ext}$, $\frac{\partial n}{\partial t}+\nabla \cdot
\mathbf{j}=0 $ and $\frac{\partial (ns)}{\partial t}+\nabla \cdot (ns
\mathbf{v}_{n})=0$, with $\Pi^{ij}=p \delta^{ij}+n_{s}
\mathbf{v}_{s}^{i} \mathbf{v}_{s}^{j}+n_{n} \mathbf{v}_{n}^{i}
\mathbf{v}_{n}^{j}$, $n=n_{s}+n_{n}$ and $\mathbf{j}=n_{s}
\mathbf{v}_{s}+n_{n} \mathbf{v}_{n}$.  Here $\mu$ is the chemical
potential, $p$ the pressure, and $s$ the entropy \emph{per particle}.
Moreover, we have $n_n \mathbf{v}_n + n_s \mathbf{v}_s = n \mathbf{v}$.
Here $n_s$ ($v_s$) and $n_n$ ($v_n$) denote the superfluid and normal
densities (velocities), respectively.
We use the subscript ``0'' to denote equilibrium quantities such that
$\mathbf{v}_{s0}=\mathbf{v}_{n0}=0$, $\nabla(\mu+V_{ext})=0$, $\nabla
p_{0}=-n_{0} \nabla V_{ext}$, and $n_0$, $s_0$, $\mu_0$, $p_0$ are
independent of time $t$.  Combining this with the thermodynamic relation
$d\mu=- s\,dT+dp/n$, we have $\nabla T_0=0$, implying that temperature
$T_0$ is constant in the trap. It then follows that in equilibrium
$\mu=\mu_0-V_{ext}$, consistent with the Thomas-Fermi approximation.

For small deviations from equilibrium, we may linearize the two fluid
equations.
%
%\begin{eqnarray}
%&&m\frac{\partial \mathbf{v}_{s}}{\partial t}+\nabla \, \delta \mu=0
%\nonumber\\ 
%&&\frac{\partial \mathbf{j}}{\partial t}+\nabla \, \delta p =-\delta n
%\, \nabla V_{ext}\nonumber\\ 
%&&\frac{\partial \, \delta n}{\partial t}+\nabla \cdot \mathbf{j}=0
%\nonumber\\ 
%&&\rho_0\frac{\partial \, \delta s}{\partial
%t}+s_0\frac{\partial\,\delta n}{\partial t}+\nabla \cdot (n_0 s_0
%\mathbf{v}_{n})=0\nonumber
%\end{eqnarray}
%with $\mathbf{j}=\rho_{s0} \mathbf{v}_{s}+\rho_{n0} \mathbf{v}_{n}$
%
Eliminating the velocities $\mathbf{v}_s$ and $\mathbf{v}_n$,
one finds \cite{HoShenoy,TaylorGriffin}
\begin{equation}
\label{twoflu1}
m\frac{\p^{2} \delta n}{\p
  t^{2}}=\nabla\cdot\left(n_0\nabla\frac{\delta p}{n_0}\right)
- \nabla \cdot (\delta T \,n_0 \nabla s_0) 
\end{equation}
\begin{eqnarray}
\label{twoflu2}
m\frac{\p^{2} \delta s}{\p t^{2}} &=&
\frac{1}{n_{0}}\,\nabla\left(\frac{n_{s0} n_{0} 
    s_0^2}{n_{n0}}\nabla \delta T\right)\nonumber\\
&&{}-(\nabla s_0)^2\delta T+\nabla
s_0\cdot\nabla\left(\frac{\delta p}{n_0}\right)
\end{eqnarray}
%\begin{eqnarray}
%\label{twoflu1}
%m\frac{\p^{2} \delta n}{\p
%  t^{2}}&=&\nabla\cdot\left(n_0\nabla\frac{\delta p}{n_0}\right)
%- \nabla \cdot (\delta T \,n_0 \nabla s_0)\,, \\
%\label{twoflu2}
%m\frac{\p^{2} \delta s}{\p
%  t^{2}}&=&\frac{1}{n_{0}}\,\nabla\left(\frac{n_{s0} n_{0} 
%    s_0^2}{n_{n0}}\nabla \delta T\right)\nonumber\\
%&&{}-(\nabla s_0)^2\delta T+\nabla
%s_0\cdot\nabla\left(\frac{\delta p}{n_0}\right)\,.
%\end{eqnarray}
%
%
We will focus on $\delta\mu(r)$ and $\delta T(r)$ as the principal
variables.  This choice, which is different from that in
Ref.~\cite{HoShenoy}, is made because both variables are non vanishing
at the trap edge so that in a basis set expansion they will satisfy the
same boundary conditions.  Moreover, the two fluid equations are
simplest in this form.  Expressing $\delta s$, $\delta p$ and $\delta n$
in terms of $\delta\mu$, $\delta T$, the two fluid equations can be
rewritten as
\begin{eqnarray}
\label{muT1}
\left(\frac{\p n}{\p\mu}\right)_T \frac{\p^2\delta\mu}{\p
t^2}+\left(\frac{\p n}{\p T}\right)_\mu \frac{\p^2\delta T}{\p
t^2}&=&\frac{A}{m}\,,\\ 
\label{muT2}
\Big(\frac{\p s}{\p\mu}\Big)_T \frac{\p^2\delta\mu}{\p
  t^2}+\Big(\frac{\p s}{\p T}\Big)_\mu\frac{\p^2\delta T}{\p
  t^2}&=&\frac{B}{m}\,,
\end{eqnarray}
with
\begin{eqnarray}
A&=&\nabla\cdot(n\nabla\delta\mu)+\nabla\cdot (ns\nabla\delta T)\,,\nonumber\\
B&=&\frac{1}{n}\nabla\cdot \Big(\frac{n n_s}{n_n}s^2\nabla\delta
T\Big)
+\nabla s\cdot\nabla\delta\mu+s\nabla s\cdot\nabla\delta
T\,.\nonumber
\end{eqnarray}
It is understood that all coefficients of $\delta \mu$ and $\delta T$
are calculated in equilibrium so that we drop the subscript ``0''. The
thermodynamical quantities in equilibrium can be calculated following
Ref.~\cite{ChenThermo}, based on the standard local density
approximation (LDA), $\mu(r)=\mu_0-V_{ext}(r)$. Their derivatives with
respect to $T$ and $\mu$ can be calculated analytically, and their
gradients can be obtained via $\nabla f=-\Big(\frac{\p f}{\p\mu}\Big)_T
\nabla V_{ext}$, where $f$ denotes any of the variables ($n, n_s, n_n,
s$).

To solve the two coupled differential equations (\ref{muT1}) and
(\ref{muT2}), we assume a simple harmonic time dependence $\delta\mu,
\delta T\propto e^{-i\omega t}$. We cast the differential two fluid
equations into an eigenfunction problem with $\omega^2$ playing the role
of eigenvalue and the eigenfunctions given by the amplitudes of $\delta
\mu$ and $\delta T$.  Since neither $T$ nor $\mu$ depends on the
density, they will not vanish at the trap edge.  Our boundary conditions
require that all thermodynamic variables be smooth (but not necessarily
zero) at the trap edge. At finite $T$, the density in the trap decreases
exponentially when the local chemical potential becomes negative at
large radius.  We choose, thus, to expand $\delta\mu$ and $\delta T$ in
terms of Jacobi polynomials.  For our numerics we choose the matrix
dimension to be $300$; we have similarly investigated matrices of
dimension $200$ up to $900$, and found little change in our principal
findings.

We now turn to an important aspect of our numerics.  Because we generate
some $300$ frequencies in our numerical approach it is essential to
establish a mechanism for systematically identifying first and second
sound modes.  To help find such a ``fingerprint'', we introduce a
``decoupling approximation'' based on reducing Eqs. (\ref{muT1}) and
(\ref{muT2}) to $\ddot{\delta\mu}=g_1(\delta\mu,\delta T)$,
$\ddot{\delta T}=g_2(\delta\mu, \delta T)$, where $g_{1,2}$ are known
functions.  We eliminate cross terms by setting $\delta T =0$ in $g_1$
and $\delta \mu =0 $ in $g_2$.  With these two decoupled equations it is
then relatively straightforward to associate a profile plot of the
numerically calculated $\delta p(r)$ and $n\delta s(r)$ vs $r$ within a
trap with first or second sound-like modes.  Here we convert $\delta\mu$
and $\delta T$ to $\delta p$ and $\delta s$ via $\delta
p=n\delta\mu+ns\delta T$ and $\delta s=\Big(\frac{\p
  s}{\p\mu}\Big)_{T}\delta\mu+\Big(\frac{\p s}{\p T}\Big)_{\mu}\delta
T$.  We stress that this procedure differs somewhat from that in
Ref.~\cite{HoShenoy} where it was argued that one could associate second
sound with a mode in which the oscillations in the thermodynamical
variable $\delta T(r)$ were constrained to be in the condensate region
of the trap.  As explained below, we do not find this to be the case. An
important check on our procedure is that we find that there is no sign
of second sound above $T_c$.

Up to this point everything is general applying to both Fermi and Bose
superfluids.  All that is needed is a microscopic theory for
thermodynamical variables. Here we use a calculational framework we have
developed for treating BCS-BEC crossover in trapped Fermi gases
\cite{ChenThermo,JS5} which emphasizes the importance of pseudogap
effects or finite momentum pairs.  The local thermodynamical potential
(density) $\Omega=\Omega_f+\Omega_b$ is associated with a contribution
from gapped fermionic excitations $\Omega_f$ as well as from
non-condensed pairs, called $\Omega_b$. We have
\begin{eqnarray}
\Omega_f&=&-\frac{\Delta^2}{g}+\sumk[(\xik-\Ek)-2T\ln(1+e^{-\Ek/T})],\nonumber\\ 
\Omega_b&=&-Z\Delta^2\mu_{pair}+\sumq T\ln(1-e^{-\Omegaq/T})\,.
\end{eqnarray}
Here $\mu_{pair}$ is the chemical potential of the pairs which is zero
below $T_c$ and the pair dispersion $\Omegaq$, along with the (inverse)
residue $Z$, can be derived from a microscopic $T$-matrix theory,
described elsewhere \cite{ChenPRL98,ourreview}.  Using $\Omega$ one then
arrives at thermodynamical properties such as the entropy density $
ns=-\frac{\p\Omega}{\p T}$ as well as self consistent equations for the
total excitation gap $\Delta$, the contribution to $\Delta$ from
noncondensed pairs (called the pseudogap) and the number equations.
These self-consistent (local) equations are simply given by
$\frac{\p\Omega}{\p\Delta}=0$, $\frac{\p\Omega}{\p\mu_{pair}}=0$ and
$n=-\frac{\p\Omega}{\p\mu}$, subject to the total number constraint
$N=\int d^3 r\, n(r)$.
When $ T < T_c$, we can use the gap equation and chain rule to
eliminate the variable $\Delta$, which is a function of $\mu$ and $T$, via 
%Thus for general $f=n, n_s, s$ the
%necessary derivatives can be calculated via $ \Big(\frac{\p
%  f}{\p\mu}\Big)_T=\Big(\frac{\p f}{\p\mu}\Big)_{\Delta,T}+\Big(\frac{\p
%  f}{\p\Delta}\Big)_{\mu,T} \frac{\p\Delta}{\p\mu}$ and $\Big(\frac{\p
%  f}{\p T}\Big)_{\mu}=\Big(\frac{\p f}{\p T}\Big)_{\mu,T}+\Big(\frac{\p
%  f}{\p\Delta}\Big)_{\mu,T} \frac{\p\Delta}{\p T} $ with
$\frac{\p\Delta}{\p\mu} =-\frac{\p^2\Omega_f}{\p\Delta\p\mu}/
\frac{\p^2\Omega_f}{\p\Delta^2}$ and $\frac{\p\Delta}{\p
  T} =-\frac{\p^2\Omega_f}{\p\Delta\p
  T}/\frac{\p^2\Omega_f}{\p\Delta^2}$.
Similarly, when $T>T_c$ we use the gap and pseudogap equations to
eliminate the variables $\Delta$ and $\mu_{pair}$ to arrive at
thermodynamical quantities.

\begin{figure}
%\centerline{\includegraphics[clip,width=3.3in] {decouple.eps}}
\centerline{\includegraphics[clip,width=3.3in] {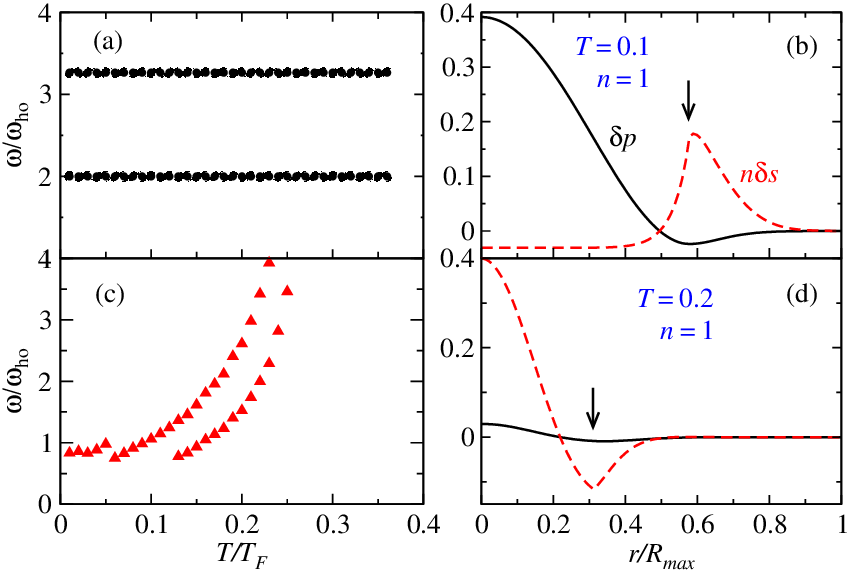}}
\caption{(Color online) Behavior of the first (upper row) and second
  sound (lower row) modes for a spherical trap at unitarity within the
  ``decoupling approximation'' (see text). The left column shows the
  $T$-dependence of the frequencies, while the right shows corresponding
  typical spatial oscillation profiles for $\delta p$ [(black) solid
  lines] and $n\delta s$ [(red) dashed lines] at the lowest frequencies,
  which provide fingerprints of first and second sound. For the second
  sound, $n\delta s$ dominates and changes sign within the condensate.
  Here $T_c \approx 0.27T_F$, and the arrows indicate the condensate
  edge.}
\label{fig:1}
\end{figure}

Figure 1 shows the lowest two collective modes at unitarity in a
spherical trap, obtained by solving the chemical potential or
temperature fluctuation equation in the decoupling approximation scheme
described earlier.  It is evident that our approximated breathing mode
frequency [Fig.~1(a)] is independent of temperature.  We understand this
result by noting that the decoupled equation for the breathing mode is
given by $-\omega^2\delta\mu=C_{\mu\mu1}\nabla^2\delta\mu
+C_{\mu\mu2}\cdot\nabla\delta\mu$ where
$C_{\mu\mu1}=n\Big(\frac{\p\mu}{\p n}\Big)_s$ and
$C_{\mu\mu2}=\nabla\mu$.  At unitarity, $n\Big(\frac{\p\mu}{\p
  n}\Big)_s=\frac{2}{3}\mu$. The only $T$-dependence contained in
$\mu_0\equiv \mu(r=0)$ can be eliminated via a simple re-scaling of $r
\rightarrow r \sqrt{2\mu_0/m\omega_{ho}^2}$, yielding a $T$-independent
breathing mode frequency.  These arguments can be shown to be equivalent
to the isentropic assumption of Ref.~\cite{ThomasUnitary}, where
$\mathbf{v}_s=\mathbf{v}_n$ is assumed in the two fluid model, leading
to two simpler coupled equations associated with the Euler and the
continuity equations.

By contrast, the second sound mode frequency we obtain increases rapidly with
temperature.  Some typical oscillation profiles of $\delta p(r)$ and
$n\delta s(r)$ are shown in the right two panels of Figure 1.  Although
the ``entropy density oscillations'' $n\delta s(r)$ fall off at large
$r$, the entropy per particle $\delta s(r)$ oscillations (not shown)
increase very rapidly upon entering the normal region.
Consequently temperature fluctuations $\delta T$ become large at the
trap edge.

Our identification of first sound for the decoupled case leads us to
associate this mode in a coupled situation with a profile for which at
the trap center $\delta p$ has a large amplitude, while $n\delta s$ is
almost zero (with a small peak near the trap edge) as in Fig.~1(b).  By
contrast, in the trap center the second sound mode has large entropy
fluctuations, while the pressure fluctuations are almost zero
[Fig.~1(d)].  These features will serve as fingerprints for
distinguishing (lower order) first and second sound modes at finite
temperatures.

\begin{figure}
%\centerline{\includegraphics[clip,width=3.3in]{eigenfunc.eps}}
\centerline{\includegraphics[clip,width=3.3in]{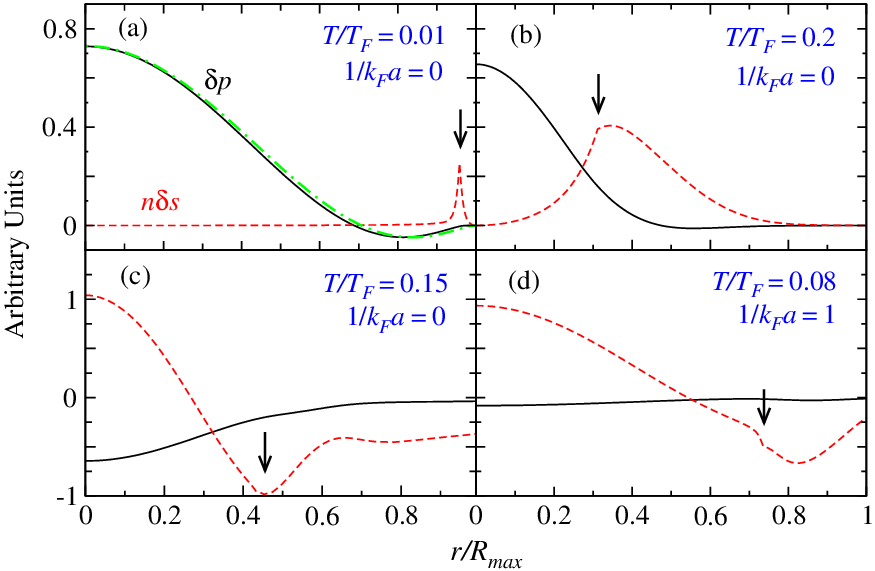}}
\caption{(Color online) Typical spatial oscillation profiles for $\delta
  p$ [(black) solid lines] and $n\delta s$ [(red) dashed lines] obtained
  from the fully coupled equations for a spherical trap at (a-c)
  unitarity and (d) $1/k_Fa=1$ at different $T$, for the first (top row)
  and second (bottom row) sound modes. Also shown in (a) is the $T=0$
  analytical result [(green) dot-dashed curve].  The arrows indicate
  condensate edge. }
\label{fig:2}
\end{figure}

Figure 2 shows some typical eigenfunction profiles of the lowest modes
obtained in the spherical trap upon solution of the \textit{fully coupled} two
fluid equations.  The first row corresponds to the breathing mode in the
unitary case. The good agreement between the very low $T$ result for
$\delta p$ and the $T=0$ analytical solution (green dashed line) in
Fig.~2(a) helps validate our numerical scheme.
Figure 2(b) corresponds to a high temperature breathing mode. In this
regime the pseudogap region outside the superfluid core is relatively
large and the peak in $n\delta s(r)$ is accordingly very broad. The
lower two panels correspond to second sound modes for the unitary and
BEC cases. By contrast with the breathing modes, here $n\delta s$ has a
larger amplitude than $\delta p$ with an opposite sign.  Clearly this is
very similar to what we observed in the decoupled mode analysis of
Fig.~1.

\begin{figure}
%\centerline{\includegraphics[clip,width=3.4in] {breathonly.eps}}
\centerline{\includegraphics[clip,width=3.4in] {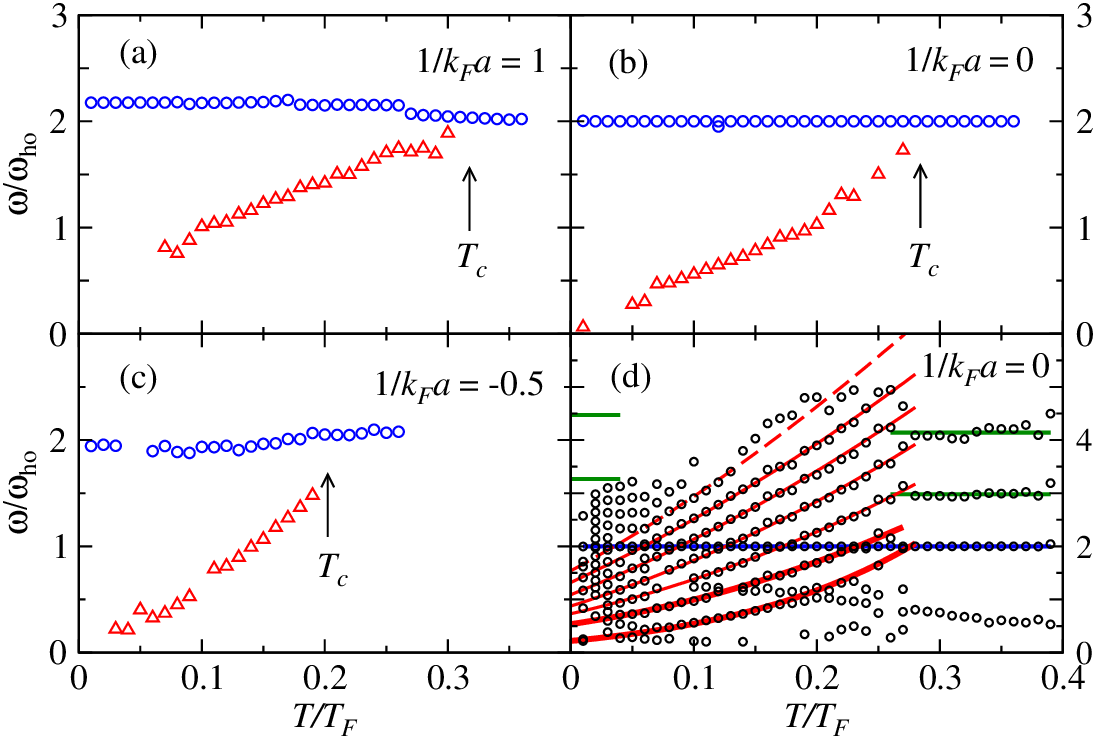}}
\caption{(Color online) Temperature dependence of breathing mode and
  second sound frequencies in a spherical trap. Panels (a-c) are for the
  near-BEC, unitary, and near-BCS cases, respectively. The upper and
  lower branches in (a-c) represent the lowest breathing mode and second
  sound frequencies, respectively. In (d), more complete results (open
  circles) are shown at unitarity, where the lines serve as guides to
  the eye for the breathing mode (nearly horizontal blue and green
  lines) and second sound (increasing red curves) frequencies.  }
\label{fig:3}
\end{figure}

Figures 3(a-c) address the fully coupled equations and show the behavior
of the lowest breathing (upper branch) and second sound mode (lower
branch) frequencies as a function of temperature in a spherical trap for
$1/k_F a=1$, 0, and $-0.5$, respectively.  For all three values of
$1/k_F a$ we find very little sign of $T_c$ in the lowest breathing mode
frequency.  In contrast, the second sound mode frequencies increase with
$T$ and disappear above $T_c$.  Figure 3(d) presents a more complete
series of modes for the unitary case.  Here the lines serve as guides to
the eye for the lowest (blue) and higher order (green) breathing mode,
and second sound (red) frequencies.

At unitarity [Fig.~3(d)] one can identify a sequence of higher order
breathing modes which precisely overlap with analytical calculations for
$T=0$. Importantly, only the lowest of these is found to be a constant
in temperature; the others are found to mix with second sound modes, as
indicated in the solid and dashed (red) lines in Fig. 3(d).  The
behavior of the lowest mode helps to justify the isentropic assumption
made in Ref.~\cite{ThomasUnitary}. We understand this by referring back
to the decoupled profiles in Figs.~1(b) and 1(c), which are seen to be
quite distinct. By contrast the profiles of the decoupled first and
second sound modes at higher order (not shown) appear more similar to
each other than their lowest order counterparts. Indeed, the behavior of
the profiles at higher order is associated with an increasing number of
nodes in the curves of $\delta p$ and $n \delta s$ (not shown), which
leads to a greater similarity between first and second sound profiles
and helps explain why the higher order modes are more readily mixed
\cite{lowlyingmode}.

In summary, we have presented predictions for future experiments on
higher breathing modes and second sound in a trap.  We find that
\textit{only the lowest} breathing mode frequency has very weak $T$
dependence.  For the unitary case this temperature insensitivity was
clearly observed by Thomas and co-workers \cite{ThomasUnitary}.  As a
result of this experiment it should not be surprising that we find
relatively weak dependencies on either side of the Feshbach resonance
for this breathing mode frequency.  In the literature there are
experimental claims (at $1/k_Fa = 1.0$) which are consistent with a
decrease \cite{GrimmCollective} in the radial breathing mode frequency,
as indeed we find here, although ours is probably too weak to reconcile
the different findings in Refs.~\cite{GrimmCollective} and
\cite{Thomas2,Kinast2}.
Finally, our more complete studies at unitarity show that if higher
order radial breathing modes can be accessed, because of their strong
hybridization with second sound, it may be possible to use these
breathing modes, rather than direct second sound to determine the
transition temperatures $T_c$.  This should be of value since there are
currently few experiments which can assign a value to $T_c$ over the
wider crossover regime.

This work was supported by Grant Nos. NSF PHY-0555325 and NSF-MRSEC
DMR-0213745; we thank John Thomas for suggesting this problem and Cheng
Chin for useful conversations.

\bibliographystyle{apsrev}

%\bibliography{Review2}

\begin{thebibliography}{25}
\expandafter\ifx\csname natexlab\endcsname\relax\def\natexlab#1{#1}\fi
\expandafter\ifx\csname bibnamefont\endcsname\relax
  \def\bibnamefont#1{#1}\fi
\expandafter\ifx\csname bibfnamefont\endcsname\relax
  \def\bibfnamefont#1{#1}\fi
\expandafter\ifx\csname citenamefont\endcsname\relax
  \def\citenamefont#1{#1}\fi
\expandafter\ifx\csname url\endcsname\relax
  \def\url#1{\texttt{#1}}\fi
\expandafter\ifx\csname urlprefix\endcsname\relax\def\urlprefix{URL }\fi
\providecommand{\bibinfo}[2]{#2}
\providecommand{\eprint}[2][]{\url{#2}}

\bibitem[{\citenamefont{Chin et~al.}(2004)\citenamefont{Chin, Bartenstein,
  Altmeyer, Riedl, Jochim, Hecker-Denschlag, and Grimm}}]{Grimm4}
\bibinfo{author}{\bibfnamefont{C.}~\bibnamefont{Chin}},
  \bibinfo{author}{\bibfnamefont{M.}~\bibnamefont{Bartenstein}},
  \bibinfo{author}{\bibfnamefont{A.}~\bibnamefont{Altmeyer}},
  \bibinfo{author}{\bibfnamefont{S.}~\bibnamefont{Riedl}},
  \bibinfo{author}{\bibfnamefont{S.}~\bibnamefont{Jochim}},
  \bibinfo{author}{\bibfnamefont{J.}~\bibnamefont{Hecker-Denschlag}},
  \bibnamefont{and} \bibinfo{author}{\bibfnamefont{R.}~\bibnamefont{Grimm}},
  \bibinfo{journal}{Science} \textbf{\bibinfo{volume}{305}},
  \bibinfo{pages}{1128} (\bibinfo{year}{2004}).

\bibitem[{\citenamefont{Kinast et~al.}(2004{\natexlab{a}})\citenamefont{Kinast,
  Hemmer, Gehm, Turlapov, and Thomas}}]{Thomas2}
\bibinfo{author}{\bibfnamefont{J.}~\bibnamefont{Kinast}},
  \bibinfo{author}{\bibfnamefont{S.~L.} \bibnamefont{Hemmer}},
  \bibinfo{author}{\bibfnamefont{M.~E.} \bibnamefont{Gehm}},
  \bibinfo{author}{\bibfnamefont{A.}~\bibnamefont{Turlapov}}, \bibnamefont{and}
  \bibinfo{author}{\bibfnamefont{J.~E.} \bibnamefont{Thomas}},
  \bibinfo{journal}{Phys. Rev. Lett.} \textbf{\bibinfo{volume}{92}},
  \bibinfo{pages}{150402} (\bibinfo{year}{2004}{\natexlab{a}}).

\bibitem[{\citenamefont{Kinast et~al.}(2004{\natexlab{b}})\citenamefont{Kinast,
  Turlapov, and Thomas}}]{Kinast2}
\bibinfo{author}{\bibfnamefont{J.}~\bibnamefont{Kinast}},
  \bibinfo{author}{\bibfnamefont{A.}~\bibnamefont{Turlapov}}, \bibnamefont{and}
  \bibinfo{author}{\bibfnamefont{J.~E.} \bibnamefont{Thomas}},
  \bibinfo{journal}{Phys. Rev. A} \textbf{\bibinfo{volume}{70}},
  \bibinfo{pages}{051401(R)} (\bibinfo{year}{2004}{\natexlab{b}}).

\bibitem[{\citenamefont{Altmeyer et~al.}(2007)\citenamefont{Altmeyer, Riedl,
  Kohstall, Wright, Geursen, Bartenstein, Chin, Denschlag, and
  Grimm}}]{GrimmCollective}
\bibinfo{author}{\bibfnamefont{A.}~\bibnamefont{Altmeyer}},
  \bibinfo{author}{\bibfnamefont{S.}~\bibnamefont{Riedl}},
  \bibinfo{author}{\bibfnamefont{C.}~\bibnamefont{Kohstall}},
  \bibinfo{author}{\bibfnamefont{M.~J.} \bibnamefont{Wright}},
  \bibinfo{author}{\bibfnamefont{R.}~\bibnamefont{Geursen}},
  \bibinfo{author}{\bibfnamefont{M.}~\bibnamefont{Bartenstein}},
  \bibinfo{author}{\bibfnamefont{C.}~\bibnamefont{Chin}},
  \bibinfo{author}{\bibfnamefont{J.~H.} \bibnamefont{Denschlag}},
  \bibnamefont{and} \bibinfo{author}{\bibfnamefont{R.}~\bibnamefont{Grimm}},
  \bibinfo{journal}{Phys. Rev. Lett.} \textbf{\bibinfo{volume}{98}},
  \bibinfo{pages}{040401} (\bibinfo{year}{2007}).

\bibitem[{\citenamefont{Kinast et~al.}(2005{\natexlab{a}})\citenamefont{Kinast,
  Turlapov, and Thomas}}]{Thomasnew}
\bibinfo{author}{\bibfnamefont{J.}~\bibnamefont{Kinast}},
  \bibinfo{author}{\bibfnamefont{A.}~\bibnamefont{Turlapov}}, \bibnamefont{and}
  \bibinfo{author}{\bibfnamefont{J.~E.} \bibnamefont{Thomas}},
  \bibinfo{journal}{Phys. Rev. Lett.} \textbf{\bibinfo{volume}{94}},
  \bibinfo{pages}{170404} (\bibinfo{year}{2005}{\natexlab{a}}).

\bibitem[{\citenamefont{Thomas et~al.}(2005)\citenamefont{Thomas, Kinast, and
  Turlapov}}]{ThomasUnitary}
\bibinfo{author}{\bibfnamefont{J.~E.} \bibnamefont{Thomas}},
  \bibinfo{author}{\bibfnamefont{J.}~\bibnamefont{Kinast}}, \bibnamefont{and}
  \bibinfo{author}{\bibfnamefont{A.}~\bibnamefont{Turlapov}},
  \bibinfo{journal}{Phys. Rev. Lett.} \textbf{\bibinfo{volume}{95}},
  \bibinfo{pages}{120402} (\bibinfo{year}{2005}).

\bibitem[{\citenamefont{Joseph et~al.}(2006)\citenamefont{Joseph, Clancy, Luo,
  Kinast, Turlapov, and Thomas}}]{ThomasSound}
\bibinfo{author}{\bibfnamefont{J.}~\bibnamefont{Joseph}},
  \bibinfo{author}{\bibfnamefont{B.}~\bibnamefont{Clancy}},
  \bibinfo{author}{\bibfnamefont{L.}~\bibnamefont{Luo}},
  \bibinfo{author}{\bibfnamefont{J.}~\bibnamefont{Kinast}},
  \bibinfo{author}{\bibfnamefont{A.}~\bibnamefont{Turlapov}}, \bibnamefont{and}
  \bibinfo{author}{\bibfnamefont{J.~E.} \bibnamefont{Thomas}}
  (\bibinfo{year}{2006}), \bibinfo{note}{cond-mat/0612567}.

\bibitem[{\citenamefont{Hu et~al.}(2004)\citenamefont{Hu, Minguzzi, Liu, and
  Tosi}}]{Tosi}
\bibinfo{author}{\bibfnamefont{H.}~\bibnamefont{Hu}},
  \bibinfo{author}{\bibfnamefont{A.}~\bibnamefont{Minguzzi}},
  \bibinfo{author}{\bibfnamefont{X.-J.} \bibnamefont{Liu}}, \bibnamefont{and}
  \bibinfo{author}{\bibfnamefont{M.~P.} \bibnamefont{Tosi}},
  \bibinfo{journal}{Phys. Rev. Lett.} \textbf{\bibinfo{volume}{93}},
  \bibinfo{pages}{190403} (\bibinfo{year}{2004}).

\bibitem[{\citenamefont{Heiselberg}(2004)}]{Heiselberg}
\bibinfo{author}{\bibfnamefont{H.}~\bibnamefont{Heiselberg}},
  \bibinfo{journal}{Phys. Rev. Lett.} \textbf{\bibinfo{volume}{93}},
  \bibinfo{pages}{040402} (\bibinfo{year}{2004}).

\bibitem[{\citenamefont{Astrakharchik et~al.}(2005)\citenamefont{Astrakharchik,
  Combescot, Leyronas, and Stringari}}]{MCCollective}
\bibinfo{author}{\bibfnamefont{G.~E.} \bibnamefont{Astrakharchik}},
  \bibinfo{author}{\bibfnamefont{R.}~\bibnamefont{Combescot}},
  \bibinfo{author}{\bibfnamefont{X.}~\bibnamefont{Leyronas}}, \bibnamefont{and}
  \bibinfo{author}{\bibfnamefont{S.}~\bibnamefont{Stringari}},
  \bibinfo{journal}{Phys. Rev. Lett.} \textbf{\bibinfo{volume}{95}},
  \bibinfo{pages}{030404} (\bibinfo{year}{2005}).

\bibitem[{\citenamefont{Griffin and Zaremba}(1997)}]{GriffinZaremba}
\bibinfo{author}{\bibfnamefont{A.}~\bibnamefont{Griffin}} \bibnamefont{and}
  \bibinfo{author}{\bibfnamefont{E.}~\bibnamefont{Zaremba}},
  \bibinfo{journal}{Phys. Rev. A} \textbf{\bibinfo{volume}{56}},
  \bibinfo{pages}{4839} (\bibinfo{year}{1997}).

\bibitem[{\citenamefont{Taylor and Griffin}(2005)}]{TaylorGriffin}
\bibinfo{author}{\bibfnamefont{E.}~\bibnamefont{Taylor}} \bibnamefont{and}
  \bibinfo{author}{\bibfnamefont{A.}~\bibnamefont{Griffin}},
  \bibinfo{journal}{Phys. Rev. A} \textbf{\bibinfo{volume}{72}},
  \bibinfo{pages}{053630} (\bibinfo{year}{2005}).

\bibitem[{\citenamefont{Chen et~al.}(1998)\citenamefont{Chen, Kosztin, Jank\'o,
  and Levin}}]{ChenPRL98}
\bibinfo{author}{\bibfnamefont{Q.~J.} \bibnamefont{Chen}},
  \bibinfo{author}{\bibfnamefont{I.}~\bibnamefont{Kosztin}},
  \bibinfo{author}{\bibfnamefont{B.}~\bibnamefont{Jank\'o}}, \bibnamefont{and}
  \bibinfo{author}{\bibfnamefont{K.}~\bibnamefont{Levin}},
  \bibinfo{journal}{Phys. Rev. Lett.} \textbf{\bibinfo{volume}{81}},
  \bibinfo{pages}{4708} (\bibinfo{year}{1998}).

\bibitem[{\citenamefont{Chen et~al.}(2005{\natexlab{a}})\citenamefont{Chen,
  Stajic, Tan, and Levin}}]{ourreview}
\bibinfo{author}{\bibfnamefont{Q.~J.} \bibnamefont{Chen}},
  \bibinfo{author}{\bibfnamefont{J.}~\bibnamefont{Stajic}},
  \bibinfo{author}{\bibfnamefont{S.~N.} \bibnamefont{Tan}}, \bibnamefont{and}
  \bibinfo{author}{\bibfnamefont{K.}~\bibnamefont{Levin}},
  \bibinfo{journal}{Phys. Rep.} \textbf{\bibinfo{volume}{412}},
  \bibinfo{pages}{1} (\bibinfo{year}{2005}{\natexlab{a}}).

\bibitem[{\citenamefont{Zwierlein et~al.}(2005)\citenamefont{Zwierlein,
  Abo-Shaeer, Schirotzek, and Ketterle}}]{KetterleV}
\bibinfo{author}{\bibfnamefont{M.~W.} \bibnamefont{Zwierlein}},
  \bibinfo{author}{\bibfnamefont{J.~R.} \bibnamefont{Abo-Shaeer}},
  \bibinfo{author}{\bibfnamefont{A.}~\bibnamefont{Schirotzek}},
  \bibnamefont{and} \bibinfo{author}{\bibfnamefont{W.}~\bibnamefont{Ketterle}},
  \bibinfo{journal}{Nature} \textbf{\bibinfo{volume}{435}},
  \bibinfo{pages}{170404} (\bibinfo{year}{2005}).

\bibitem[{\citenamefont{Zwierlein et~al.}(2006)\citenamefont{Zwierlein,
  Schirotzek, Schunck, and Ketterle}}]{ZSSK06}
\bibinfo{author}{\bibfnamefont{M.~W.} \bibnamefont{Zwierlein}},
  \bibinfo{author}{\bibfnamefont{A.}~\bibnamefont{Schirotzek}},
  \bibinfo{author}{\bibfnamefont{C.~H.} \bibnamefont{Schunck}},
  \bibnamefont{and} \bibinfo{author}{\bibfnamefont{W.}~\bibnamefont{Ketterle}},
  \bibinfo{journal}{Science} \textbf{\bibinfo{volume}{311}},
  \bibinfo{pages}{492} (\bibinfo{year}{2006}).

\bibitem[{\citenamefont{Kinast et~al.}(2005{\natexlab{b}})\citenamefont{Kinast,
  Turlapov, Thomas, Chen, Stajic, and Levin}}]{ThermoScience}
\bibinfo{author}{\bibfnamefont{J.}~\bibnamefont{Kinast}},
  \bibinfo{author}{\bibfnamefont{A.}~\bibnamefont{Turlapov}},
  \bibinfo{author}{\bibfnamefont{J.~E.} \bibnamefont{Thomas}},
  \bibinfo{author}{\bibfnamefont{Q.~J.} \bibnamefont{Chen}},
  \bibinfo{author}{\bibfnamefont{J.}~\bibnamefont{Stajic}}, \bibnamefont{and}
  \bibinfo{author}{\bibfnamefont{K.}~\bibnamefont{Levin}},
  \bibinfo{journal}{Science} \textbf{\bibinfo{volume}{307}},
  \bibinfo{pages}{1296} (\bibinfo{year}{2005}{\natexlab{b}}),
  \bibinfo{note}{published online 27 January 2005;
  doi:10.1126/science.1109220}.

\bibitem[{\citenamefont{Luo et~al.}(2007)\citenamefont{Luo, Clancy, Joseph,
  Kinast, and Thomas}}]{ThomasEntropy}
\bibinfo{author}{\bibfnamefont{L.}~\bibnamefont{Luo}},
  \bibinfo{author}{\bibfnamefont{B.}~\bibnamefont{Clancy}},
  \bibinfo{author}{\bibfnamefont{J.}~\bibnamefont{Joseph}},
  \bibinfo{author}{\bibfnamefont{J.}~\bibnamefont{Kinast}}, \bibnamefont{and}
  \bibinfo{author}{\bibfnamefont{J.~E.} \bibnamefont{Thomas}},
  \bibinfo{journal}{Phys. Rev. Lett.} \textbf{\bibinfo{volume}{98}},
  \bibinfo{pages}{080402} (\bibinfo{year}{2007}).

\bibitem[{\citenamefont{Stamper-Kurn et~al.}(1998)\citenamefont{Stamper-Kurn,
  Miesner, Inouye, Andrews, and Ketterle}}]{Stamper-Kurn}
\bibinfo{author}{\bibfnamefont{D.~M.} \bibnamefont{Stamper-Kurn}},
  \bibinfo{author}{\bibfnamefont{H.-J.} \bibnamefont{Miesner}},
  \bibinfo{author}{\bibfnamefont{S.}~\bibnamefont{Inouye}},
  \bibinfo{author}{\bibfnamefont{M.~R.} \bibnamefont{Andrews}},
  \bibnamefont{and} \bibinfo{author}{\bibfnamefont{W.}~\bibnamefont{Ketterle}},
  \bibinfo{journal}{Phys. Rev. Lett.} \textbf{\bibinfo{volume}{81}},
  \bibinfo{pages}{500} (\bibinfo{year}{1998}).

\bibitem[{\citenamefont{Shenoy and Ho}(1998)}]{HoShenoy}
\bibinfo{author}{\bibfnamefont{V.~B.} \bibnamefont{Shenoy}} \bibnamefont{and}
  \bibinfo{author}{\bibfnamefont{T.-L.} \bibnamefont{Ho}},
  \bibinfo{journal}{Phys. Rev. Lett.} \textbf{\bibinfo{volume}{80}},
  \bibinfo{pages}{3895} (\bibinfo{year}{1998}).

\bibitem[{\citenamefont{Griffin et~al.}(1997)\citenamefont{Griffin, Wu, and
  Stringari}}]{StringariGriffin}
\bibinfo{author}{\bibfnamefont{A.}~\bibnamefont{Griffin}},
  \bibinfo{author}{\bibfnamefont{W.-C.} \bibnamefont{Wu}}, \bibnamefont{and}
  \bibinfo{author}{\bibfnamefont{S.}~\bibnamefont{Stringari}},
  \bibinfo{journal}{Phys. Rev. Lett.} \textbf{\bibinfo{volume}{78}},
  \bibinfo{pages}{1838} (\bibinfo{year}{1997}).

\bibitem[{\citenamefont{Heiselberg}(2006)}]{Heiselberg5}
\bibinfo{author}{\bibfnamefont{H.}~\bibnamefont{Heiselberg}},
  \bibinfo{journal}{\pra} \textbf{\bibinfo{volume}{73}},
  \bibinfo{pages}{013607} (\bibinfo{year}{2006}).

\bibitem[{\citenamefont{Chen et~al.}(2005{\natexlab{b}})\citenamefont{Chen,
  Stajic, and Levin}}]{ChenThermo}
\bibinfo{author}{\bibfnamefont{Q.~J.} \bibnamefont{Chen}},
  \bibinfo{author}{\bibfnamefont{J.}~\bibnamefont{Stajic}}, \bibnamefont{and}
  \bibinfo{author}{\bibfnamefont{K.}~\bibnamefont{Levin}},
  \bibinfo{journal}{\prl} \textbf{\bibinfo{volume}{95}},
  \bibinfo{pages}{260405} (\bibinfo{year}{2005}{\natexlab{b}}).

\bibitem[{\citenamefont{Stajic et~al.}(2005)\citenamefont{Stajic, Chen, and
  Levin}}]{JS5}
\bibinfo{author}{\bibfnamefont{J.}~\bibnamefont{Stajic}},
  \bibinfo{author}{\bibfnamefont{Q.~J.} \bibnamefont{Chen}}, \bibnamefont{and}
  \bibinfo{author}{\bibfnamefont{K.}~\bibnamefont{Levin}},
  \bibinfo{journal}{Phys. Rev. Lett.} \textbf{\bibinfo{volume}{94}},
  \bibinfo{pages}{060401} (\bibinfo{year}{2005}).

\bibitem[{low()}]{lowlyingmode}
\bibinfo{note}{It is interesting to note that there exists a lower frequency
  first-sound like mode at finite $T$, which we also find to be present in the
  ``free'' (but hydrodynamic) Fermi gas limit at $T \neq 0$.}

\end{thebibliography}

\end{document}